\documentclass[conference]{IEEEtran}
\IEEEoverridecommandlockouts
\usepackage{array}
\usepackage{cite}
\usepackage{listings}
\usepackage{amsmath,amssymb,amsfonts}
\usepackage{algorithmic}
\usepackage{graphicx}
\usepackage{textcomp}
\usepackage{xcolor}
\usepackage{multirow}
\usepackage{footnote}
\usepackage[hyphenbreaks]{breakurl}

\def\BibTeX{{\rm B\kern-.05em{\sc i\kern-.025em b}\kern-.08em
    T\kern-.1667em\lower.7ex\hbox{E}\kern-.125emX}}
\begin{document}
\setcounter{footnote}{-1}
\title{PSCS: A Path-based Neural Model for Semantic Code Search}
\author{\IEEEauthorblockN{Zhensu Sun, Yan Liu\thanks{\IEEEauthorrefmark{4}Yan Liu is corresponding author.}\IEEEauthorrefmark{4}, Chen Yang, Yu Qian}
\IEEEauthorblockA{\textit{School of Software Engineering} \\
\textit{Tongji University}\\
Shanghai, China \\
\{87su, yanliu.sse, 1610833, 1831582\}@tongji.edu.cn}
}
\maketitle

\begin{abstract}
To obtain code snippets for reuse, programmers prefer to search for related documents, e.g., blogs or Q\&A, instead of code itself. The major reason is due to the semantic diversity and mismatch between queries and code snippets. Deep learning models have been proposed to address this challenge. Compared with approaches using information retrieval techniques, deep learning models do not suffer from the information loss caused by refining user intention into keywords. However, the performance of previous works is not satisfactory because they ignore the importance of code structure. When the semantics of code (e.g., identifier names, APIs) are ambiguous, code structure may be the only feature for the model to utilize. In that case, previous works relearn the structural information from lexical tokens of code, which is extremely difficult for a model without any domain knowledge. In this work, we propose PSCS, a path-based neural model for semantic code search. Our model encodes both the semantics and structures of code represented by AST paths. We train and evaluate our model over 330k-19k query-function pairs, respectively. The evaluation results demonstrate that PSCS achieves a SuccessRate of 47.6\% and a Mean Reciprocal Rank (MRR) of 30.4\% when considering the top-10 results with a match. The proposed approach significantly outperforms both DeepCS, the first approach that applies deep learning to code search task, and CARLCS, a state-of-the-art approach that introduces a co-attentive representation learning model on the basis of DeepCS. The importance of code structure is demonstrated with an ablation study on code features, which enlightens model design for further studies. 

\end{abstract}


\section{Introduction}
As an essential part of code reuse, code search is frequently performed by programmers\cite{stolee_solving_2014}. Programmers can search on various search engines to find reusable code snippets. Compared to code search over open source repositories like Github\footnote{https://github.com/} directly, programmers prefer to search for the related documents (e.g., blogs, Q\&A) using natural language.

The reason why programmers can not search code snippets effectively from large repositories by describing intentions lies in the semantic diversity between the query and the code. Following its syntax, code of particular programing language may not have common lexical tokens with queries. For example, if users want to search for a code snippet with loops, it is better to use the word \textit{for} instead of \textit{loop} as the keyword. Therefore, following the way of document retrieval to search for code can not receive ideal results. 

Many previous studies\cite{lv_codehow_2015}\cite{raghothaman_swim_2016}\cite{niu_learning_2017}\cite{zhang_expanding_2018} apply information retrieval techniques to bridge this gap, which improves the accuracy of the code search task. For example, Nie et al.\cite{nie_query_2016} expand the query using crowd knowledge from Stack Overflow. Niu et al.\cite{niu_learning_2017} use machine learning to rank the search results. Previous works focus on either a better query (e.g., query expansion) or a better result (e.g., result ranking). The only constant in previous works is the keyword matching technique. Usually, users need to refine their intention into keywords. Otherwise, the result list may be noisy. Table~\ref{comparison} lists examples of user intentions with refined technical keywords. Refining an intention into technical keywords may lost detailed information sometimes, especially for inexperienced programmers. In most cases, the keywords can generally express the user intention. But there still exists a margin. Some detailed words may be lost like the word "last" in intention 9. It will be more precise if the search engine can directly utilize the user intention.

\begin{table*}[htbp]
\caption{The comparison between intentions and keywords for code search}
\renewcommand{\arraystretch}{1.3}
\begin{center}
\begin{tabular}{ll}
\hline
\textbf{Intention}& \textbf{Keyword}\\
\hline
1. determine if a user is already authenticated & 1. authentication check\\ 
2. redirect the response to the supplied url & 2. redirect response url\\
3. perform url encoding with utf & 3. url encoding utf\\ 
4. extract a sub array of bytes out of the byte array & 4. extract sub array\\
5. resolve service ip address & 5. resolve ip\\
6. verify whether value of cookie satisfies specified matcher & 6. verify cookie \\
7. specify the hostname for the proxy & 7. hostname proxy\\
8. add a header to be sent with the request & 8. add header request \\
9. get the result of an xml path expression & 9. parse xml path expression\\
10. remove last camel word & 10. remove camel word \\

\hline
\end{tabular}
\label{comparison}
\end{center}
\end{table*}

Inspired by the emerging of deep learning, Gu et al.\cite{gu_deep_2018} propose a deep learning model called DeepCS. It learns to jointly embed the queries and code snippets into the same high-dimensional vector space, in which the similarity between queries and code snippets can be directly calculated. In this way, users can feed the full description of their intentions into the model and obtain matched code snippets based on the semantic information. The subsequent studies of DeepCS widely adopt its architecture. The experimental results of DeepCS show that it outperforms previous 
information-retrieval-based techniques.

After that, a multitude of research makes improvements using the attention mechanism that is also widely used in numerous software engineering tasks\cite{hu2018deep}\cite{sun_req2lib_2020}. UNIF, proposed by Cambronero et al.\cite{cambronero_when_2019}, extends an existing unsupervised technique by computing the importance of each code tokens with attention. Wan et al.\cite{wan_multi-modal_2019} proposed a multi-modal attention network named MMAN with triple features, including abstract syntax tree (AST) and control-flow graph (CFG). To learn the correlations between queries and code snippets, Shuai et al.\cite{shuai_improving_2020} recently introduce a co-attentive representation learning model on the basis of DeepCS.

The lack of accessible code structure limits the performance of these previous works. It is quite common that developers use the abbreviations of identifier names for temporary convenience. In that case, the model needs to reason about the code employing structural information. Thus, understanding the structure of code is necessary for code search models. However, in DeepCS\cite{gu_deep_2018} and CARLCS\cite{shuai_improving_2020}, code snippets are represented by three shallow human-made features (i.e., method names, API sequences, and tokens) where structural information hides in text. With these features, their models have to learn the structural information from lexical tokens without any domain knowledge. It is hard even for human. Imagine a novice learning the syntax by just reading the source code. The understanding learned from code tokens is insufficient.

In particular, MMAN\cite{wan_multi-modal_2019} takes structural information into account by utilizing the AST and CFG of source code. However, the results indicate that the code structure makes a very small contribution to the model. The involvement of AST and CFG only contribute to a slight increment of 0.015 in terms of recall@10, compared to the token-only variant which achieves a recall@10 of 0.619. It indicates that the AST and CFG can not play their full roles when directly fed to a tree-based neural model. The reason is that the tree-based model can not clearly recognize the code structure hidden in hybrid nodes of AST. A structure-sensitive model is needed.

In this work, we propose a deep learning model for semantic code search, named \textbf{P}ath-based \textbf{S}emantic \textbf{C}ode \textbf{S}earch (PSCS). To obtain a comprehensive view of code, we adopt an idea of tree-based embedding that extracts paths from the AST of code\cite{alon_general_2018}. As shown in Fig.~\ref{ast}, it presents both the semantics and structures of code by walking from one AST leaf to another. In this way, the model can capture both the semantic and structural information directly. Our model learns to embed the queries and AST paths into the same vector space, using a query encoder and a code encoder, respectively. The code encoder is designed to fit with the AST path, including a shared embedding with the input query, a bi-direction Long Short-Term Memory (bi-LSTM) model, and an attentive fusion layer. The attentive fusion layer is used to compute the contribution of each path and integrate the hybrid path representations into a single representation. During training, our model updates its parameters and embedding matrices by learning the correlation between the query and the code. With the trained model, we can generate embedding vectors for input queries and compute the cosine similarity with all the pre-embedded code vectors of the code corpus. The code snippets at the top-k positions of the ranked similarity list are the final results.

To evaluate the effectiveness of our model, we train and evaluate it on filtered CodeSearchNet dataset\cite{husain_codesearchnet_2019} containing {\raise.17ex\hbox{$\scriptstyle\mathtt{\sim}$}}330k and {\raise.17ex\hbox{$\scriptstyle\mathtt{\sim}$}}19k code-query pairs, respectively. The results demonstrate that PSCS can retrieve code snippets accurately with a SuccessRate@10 of 47.6\% and a Mean Reciprocal Rank (MRR) of 30.4\%. These results significantly outperform all the baseline models, DeepCS and CARLCS, which respectively obtain 22.4\% and 25.5\% in terms of MRR. We also perform several ablation studies to evaluate our model comprehensively. One of the ablation studies focuses on the code structure. To the best of our knowledge, this is the first experiment to evaluate the impact of pure code structure for semantic code search. It shows that code structure and semantics separately can only lead to poor results, but when combining them, a performance boost comes true. 

In summary, the contributions of this paper are as follows:

\begin{itemize}
  \item We introduce the AST path as the representation of code snippets and propose a structure-sensitive neural model, PSCS, for semantic code search.
  \item We evaluate our model on a large-scale dataset. Compared with state of the art, the results show significantly increased retrieval performance, getting a 19.2\% improvement in MRR. 
  \item We experimentally demonstrate the importance of code structure for semantic code search: Code structure is an impactful supplementary feature for code understanding in code search.
\end{itemize}

The rest of the paper is arranged as follows. A review and discussion on related work are in Section 2, where AST paths are introduced in Section 3. In Section 4, we present PSCS, our approach for semantic code search. We outline our experiments in Section 5 and analyse their results in Section 6. We close with our conclusions in Section 7.

\section{Related Work}
\subsection{Code Search}
Before DeepCS, most of the code search researches are based on information retrieval techniques. For instance, Sourcerer\cite{linstead_sourcerer_2009} is an infrastructure for large-scale code search based on Lucene\footnote{https://lucene.apache.org/}. SNIFF\cite{chechik_sniff_2009} performs queries over source code annotated with Java API documents. CodeHow\cite{lv_codehow_2015} expanded queries with APIs from online documents at MSDN. As mentioned in Section 1, they all suffer from the information loss caused by keyword matching. 

To address this issue, Gu et al.\cite{gu_deep_2018} propose DeepCS, which is the first to apply the deep learning model to the code search task. It learns to map a natural language query and a code snippet into the same high-dimension vector space, and compute the cosine similarity between them. This architecture is adopted by most of its following works, including our work.

After that, more deep learning models for code search are proposed. Cambronero et al.\cite{cambronero_when_2019} construct a network using attention to combine per-token embedding of the source code. MMAN, proposed by Wan et al.\cite{wan_multi-modal_2019}, considers both shallow features and structured features by applying an attention mechanism to integrate the multi-modal representations. In \cite{shuai_improving_2020}, Shuai et al. introduce a co-attention mechanism to learn the interdependent representations for the embedded code and query. Those previous models do not fully utilize the deep structural information of source code. Different from previous works, our model targets on taking full advantage of the code structure. It adopts the accessible code structure to support the understanding of the semantics in code.

In particular, some researches introduce additional code-related tasks to improve the performance of code retrieval. Chen et al.\cite{chen_neural_2018} propose a neural framework with two variational autoencoders for code search and code summarization, respectively. Similarly, Ye et al.\cite{ye_leveraging_2020} introduce additional code generation task leveraging a novel end-to-end model with dual learning. The CoaCor, proposed by Yao et al.\cite{yao_coacor_2019}, introduces a code annotation model to help code search models distinguish relevant code snippets from others. Wang et al.\cite{Wang2020TranS3AT} adopt the transformer model to integrate code summarization with code search. We believe that the model we proposed can enhance this research direction.

\subsection{Code Embedding}
Code embedding is the upstream task required by various research fields of source code. In many previous works, code can be simply modeled as token sequences\cite{loyola2017neural}, character sequences\cite{bielik_phog_nodate}, or API sequences\cite{murali2017neural}. These methods hide the structural information of code in shallow features. Thus, models need to relearn the structure of code from these features, which is hard for models without any domain knowledge.

To capture more structural information before learning, researchers consider utilizing the AST. For example, Mou et al.\cite{mou2016convolutional} apply slide windows on AST to capture features effectively in different regions. Miltiadis et al.\cite{allamanis_learning_2018} represent code with graphs where nodes are identifiers and edges are syntactic and semantic relations. Piech et al.\cite{piech2015learning} utilize tree-RNN to embed the AST. PHOG, proposed by Bielik\cite{bielik_phog_nodate}, extract syntax-based contexts by traversing the AST to identify context nodes. In this way, nodes in AST can represent the code structure. 

In \cite{alon_general_2018}, Alon et al. propose a novel representation method called AST path. In its subsequent papers, Alon et al. propose two embedding models for source code, code2vec\cite{alon_code2vec_2019} and code2seq\cite{alon_code2seq_2019}. These models demonstrate the effectiveness of representing code with the AST path. The code2seq applies a seq2seq model on multiple tasks (e.g., method name prediction). The code encoder in this work is similar to the encoder of code2seq. We make some improvements for the code search task, including shared embedding and attentive confusion.

\section{AST Path}

\begin{figure}[htbp]
\centerline{\includegraphics[width=8cm]{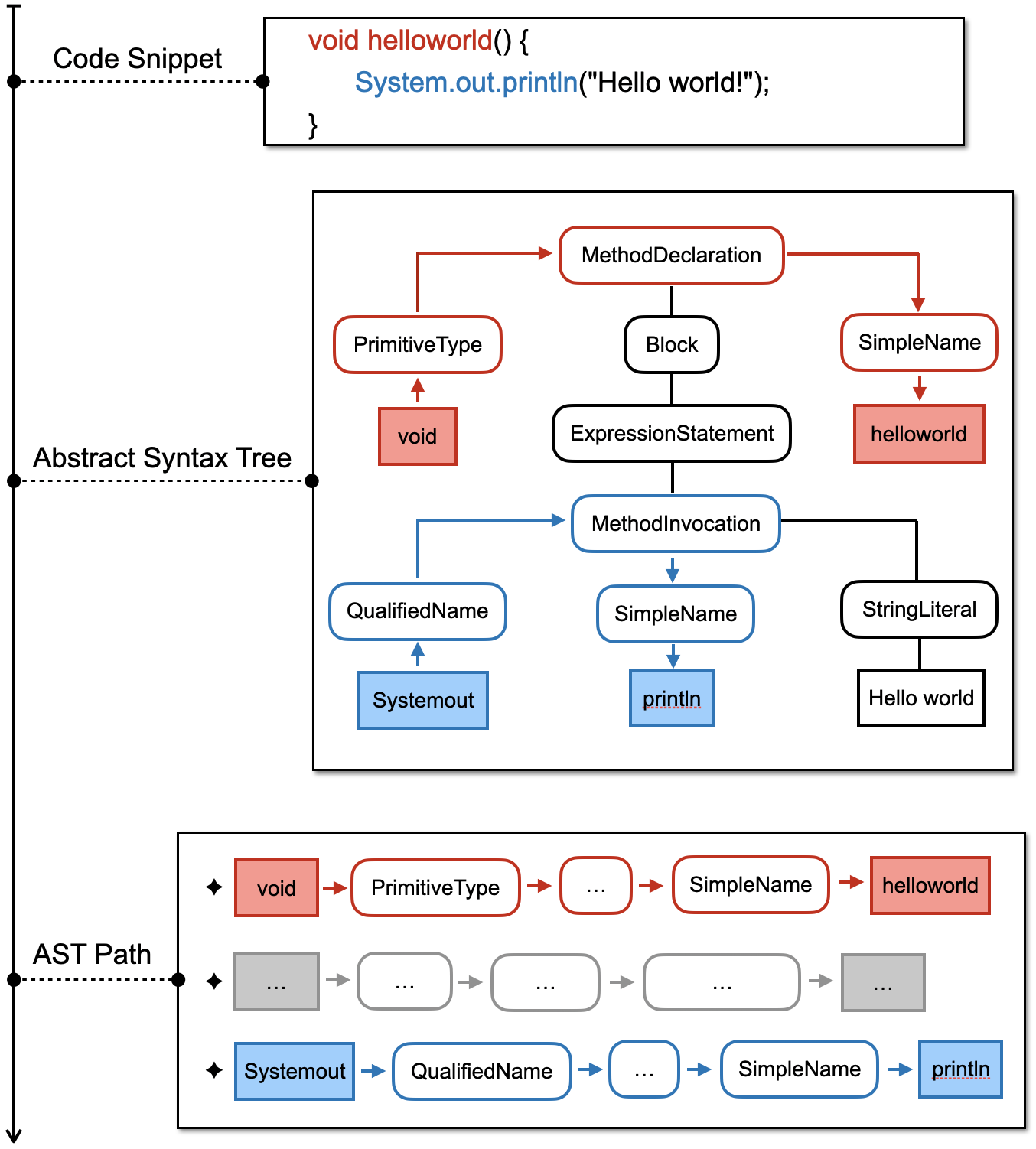}}
\caption{An example of representing a code snippet with AST paths}
\label{ast}
\end{figure}

For program languages, a code snippet can be represented by a unique abstract syntax tree (AST). An example is shown in Fig.~\ref{ast}. It is a graph consisting of two kinds of nodes: terminal nodes (i.e., the leaves) and non-terminal nodes, respectively representing user-defined identifiers (e.g., variable names) and the structure of code (e.g., a for-loop). It is widely used to present the structure of code\cite{dam2018deep}\cite{sun2019grammar}.   

The AST path is the path extracted from the AST by walking from one terminal to another. A path has non-terminal nodes in the middle and terminals at both ends. Each non-terminal node of the AST path has two types: \textit{up} and \textit{down}. The usage of different types presents the direction of paths. Here is an example path of AST in Fig.~\ref{ast}, from the function type \textit{void} to the called function \textit{println} (we abbreviate the labels of nodes): 

\begin{center}
    (void, PT↑MD↓B↓ES↓MI↓SN, prntln)
\end{center}

With AST paths, a piece of code can finally be represented by an arbitrary number of such paths. In this way, we convert the graph to euclidean data without a great loss of information. Deep learning model can utilize euclidean data efficiently. Different from shallow features, AST path contains both semantic information in terminal tokens and structural information in non-terminal nodes. It can provide a full view of code to model for a better understanding. 

\section{Proposed Model: PSCS}
\subsection{Overview}

\begin{figure*}[htbp]
\centerline{\includegraphics[width=18cm]{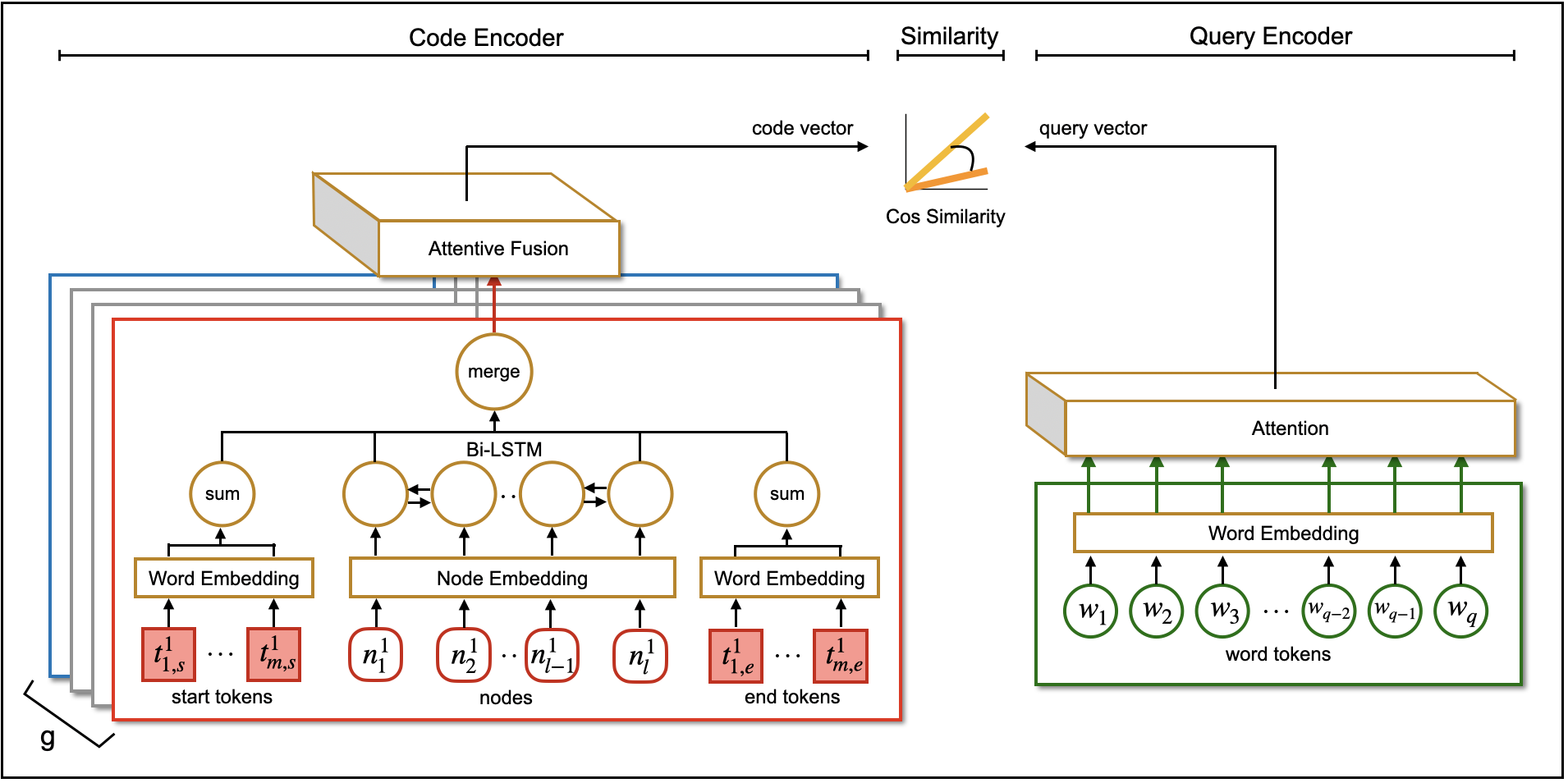}}
\caption{The overall structure of the PSCS.}
\label{overview}
\end{figure*}

In this work, we propose a deep neural model named PSCS (Path-based Semantic Code Search). To calculate the similarity between query and code, it jointly learns to encode queries and code snippets into the same vector space.  Fig.~\ref{overview} shows the overall structure of PSCS. Our model consists of a query encoder and a code encoder. We will first describe the preprocessing of data (i.e., the code-annotation pairs) and then introduce each design in detail.

\subsection{Data Preprocessing}

\subsubsection{Annotation Processing}
The first sentence of code annotation usually describes the function of corresponding code snippets using natural language. We extract the first sentence of each code annotation as the query, which is also used in the previous works\cite{gu_deep_2018}\cite{shuai_improving_2020}. After that, we remove the punctuation and bracketed texts, because these are seldom used in queries. The intentions in Table~\ref{comparison} are also the examples of extracted queries from the dataset. 
 
Some words in the sentence are camel-case or connected by hyphens. To take full use of each word, we tokenize the sentence and split each token into sub-tokens. After splitting, we convert all the sub-tokens to lower case. For example, the word \textit{setTimer} will be split into \textit{set timer}. To keep a fixed length, we pad the token sequence with a special token \textit{\textless PAD\textgreater} to the maximum length $q$. Based on the processed token corpus, we build a vocabulary that maps each word to a unique id. 

For a query $Q$, we finally obtain a sequence of its tokens $w_1w_2...w_q$, where $w$ represents the id of each word, and $q$ is the length of queries.

\subsubsection{Code Processing}
As described in Section 2, we represent code with the AST path. Typically, an AST path consists of two heterogeneous nodes: terminal nodes and non-terminal nodes. They need to be processed separately.

Usually, terminals are user-identified lexical tokens of source code expressed in natural language. Therefore, we apply the same processing steps as queries. In particular, terminals share the same vocabulary as queries for shared embedding (introduced in the next section). As for non-terminal nodes, there are no extra steps except padding and building its vocabulary.

In this way, we obtain all the paths $P$ from the corresponding code snippet of the query $Q$. The paths include the sequence of pairs of terminals $\{(t_{s}^{1},t_{e}^{1}),...,(t_{s}^{p},t_{e}^{p})\}$ and non-terminal nodes $\{(n_{1}^{1}n_{2}^{1}...n_{l}^{1}),...,(n_{1}^{p}n_{2}^{p}...n_{l}^{p})\}$, where $t$ is the sequence of terminals sub-tokens $w_1w_2...w_m$, $m$ is the maximum length of tokens, $n$ represents the id of nodes and $l$ is the maximum number of nodes in a path.

\subsection{Code Encoder}
The code encoder embeds a code snippet represented by AST paths into a high-dimensional vector. The number of paths in a code snippet is not fixed. We randomly sample $g$ paths from $P$ for each epoch during training. To integrate the information of all paths, we employ an attentive fusion layer.

\subsubsection{Terminal Token Representation}
Terminal nodes include start tokens and end tokens. There are no differences between these two tokens, except the position in the path. So the encoding methods of them are the same.

As mentioned above, we split the tokens into sub-tokens. We embeds the sub-tokens using an embedding matrix $\textbf{E}_{1} \in \mathbb{R}^{o_w \times d}$, where $o_w$ is the size of the query vocabulary, $d$ is the dimension of embedded vectors. Each row of the matrix represents the embedded vector of each token in vocabulary. The embedding matrix is initialized from a uniform distribution between 0.1 and -0.1. With this matrix, we can map the lexical tokens from ids to real value vectors. 

Finally, we sum up the embedding vectors of the sub-tokens to represent the full token:

\begin{equation}
    \textbf{e}_{start}^j = \sum_{i=1}^{m}emb_{1}(\textbf{t}_{i,s}^j)
\end{equation}

\begin{equation}
    \textbf{e}_{end}^j = \sum_{i=1}^{m}emb_{1}(\textbf{t}_{i,e}^j)
\end{equation}
where $j$ represents the $j$-th path and $emb_{1}$ is the embedding function for words.

\subsubsection{Non-terminal Node Representation}
For nodes, we apply a new embedding matrix $\textbf{E}_{2} \in \mathbb{R}^{o_n \times l}$, where $o_n$ is the size of the node vocabulary. It is initialized and utilized similarly to $\textbf{E}_{1}$. 

Different from terminal tokens, the non-terminal nodes are directional. The direction contains sequential information, which can help the model to understand the code better. Thus, we adopt a neural model called bi-direction Long Short-Term Memory (bi-LSTM)\cite{hochreiter1997long}, which can extract information in both forward and reverse directions. Compared with other neural models like Convolutional Neural Network (CNN), LSTM can capture additional sequential information. At each time step $s$, it reads the $s$-th embedded node, then computes the hidden states $\textbf{h}_s$, namely:

\begin{equation}
\overrightarrow{\textbf{h}_{s}^{j}} = \overrightarrow{LSTM}(emb_{2}(\textbf{n}_{s}^{j}), \overrightarrow{\textbf{h}_{s-1}^{j}})
\end{equation}
\begin{equation}
\overleftarrow{\textbf{h}_{s}^{j}} = \overleftarrow{LSTM}(emb_{2}(\textbf{n}_{s}^{j}), \overleftarrow{\textbf{h}_{s+1}^{j}})
\end{equation}

We concatenate the hidden states to represent non-terminal nodes:

\begin{equation}
    \textbf{e}_{node}^{j} = \overrightarrow{\textbf{h}_{s}^{j}} \oplus \overleftarrow{\textbf{h}_{s}^{j}}
\end{equation}
where $\oplus$ donates the concatenation operation.

\subsubsection{Attentive Confusion}
We combine the representation of terminal tokens and non-terminal nodes as the preliminary representation for the $j$-th path, i.e.,
\begin{equation}
    \textbf{e}_{path}^{j} = dropout(\textbf{e}_{start}^{j} \oplus \textbf{e}_{node}^{j} \oplus \textbf{e}_{end}^{j}) 
\end{equation}
where $dropout$ is the dropout regularization\cite{Srivastava2014DropoutAS} used to avoiding over-fitting by randomly dropping vector units during training. 

Among the $g$ paths, their contributions to the final representation are not equal. To dynamically measure the importance, we adopt an attention mechanism\cite{bahdanau2014neural} to compute the weighted average, where the weights for each path are computed as follows:

\begin{equation}
    \alpha_j = \frac{exp(\textbf{W}_a \textbf{e}_{path}^{j}\cdot \textbf{c}_a^T)}{\sum_{i=1}^{g}exp(\textbf{W}_a \textbf{e}_{path}^{i}\cdot \textbf{c}_a^T)}
\end{equation}

where $\textbf{W}_a$ is the matrix of trainable parameters of attention for paths, $\textbf{c}_a$ is the context vector for the code calculated by:

\begin{equation}
    \textbf{c}_a=\frac{1}{g}\sum_{j=1}^{g}\textbf{e}_{path}^{j}
\end{equation}

With the attention weights, we compute the weighted average of the path representation, then feed them to a linear layer to perform the feature-level confusion. In this way, we get the final representation vector of code: 
\begin{equation}
    \textbf{v}_{code} = \textbf{W}\sum_{j=1}^{g}\alpha_{j}\textbf{e}_{path}^{j}
\end{equation}
where $\textbf{W}$ is the weight matrix of the fully-connected layer.

\subsection{Query Encoder}
The query encoder learns the semantic information from queries using word embedding and an attention mechanism.  

For word representation, we use the same embedding matrix as the terminal tokens, which is called shared embedding, i.e., $\textbf{E}_{1} \in \mathbb{R}^{v_w \times d}$. The shared embedding is a distinctive alternative design made for the code search task. Words in the terminal nodes of AST paths are usually identified by users. They may keep the similar semantic meaning to words in queries. Sharing the embedding can save the cost of training by narrowing the margin between code representation and query representation. 

Similar to the paths, words in a sentence also make different contributions to the final representation. Thus an attention mechanism is applied to learn the weights of each word, which is calculated by:

\begin{equation}
    \beta_i = \frac{exp(\textbf{W}_b emb_{1}(\textbf{w}_i) \cdot \textbf{c}_b^T)}{\sum_{i=1}^{l}exp(\textbf{W}_bemb_{1}(\textbf{w}_i)\cdot \textbf{c}_b^T)}
\end{equation}
where $\textbf{c}_b$ is the context vector for the query, calculated by:
\begin{equation}
    \textbf{c}_b=\frac{1}{l}\sum_{i=1}^{l}emb_{1}(\textbf{w}_i)
\end{equation}

With the attention weights, we compute the weighted average of word embedding vectors to represent the whole sentence:

\begin{equation}
    \textbf{v}_{query}=\sum_{i=1}^{l}\beta_{i}emb_{1}(\textbf{w}_i)
\end{equation}

\subsection{Loss}
In this work, we adopt the pairwise ranking loss. Given a code snippet, this loss function aims to distinguish the correct query from other queries. For each query-code pair \textless$Q^+,C^+$\textgreater, we randomly select a different query $Q^-$ to construct a negative instance \textless$Q^-,C^+$\textgreater. During each epoch of training, our model encodes them into vectors and calculates the similarity for \textless$\textbf{v}_{Q}^{+},\textbf{v}_{C}^{+}$\textgreater and \textless$\textbf{v}_{Q}^{-},\textbf{v}_{C}^{+}$\textgreater. With these similarities, the model learns to minimize the loss:
\begin{equation}
    loss = \max(0,\epsilon - sim(\textbf{v}_{Q}^{+},\textbf{v}_{C}^{+}) + sim(\textbf{v}_{Q}^{-},\textbf{v}_{C}^{+}))
\end{equation}
where $\epsilon$ is a constant margin to prevent loss value from becoming zero, $\textbf{v}_{Q}$ is the query vector generated by query encoder for query $Q$, $\textbf{v}_{C}$ refers to the code vector generated by code encoder for code $C$ and $sim(\textbf{x},\textbf{y})$ refers to the function that measure the similarity between vector $x$ and $y$. In this work, $sim(\textbf{x},\textbf{y})$ represents the cosine similarity calculated by:

\begin{equation}
    sim(\textbf{x},\textbf{y}) = \dfrac{\textbf{x} \cdot \textbf{y}}{\max(\Vert \textbf{x} \Vert \cdot \Vert \textbf{y} \Vert, \delta)}
\end{equation}
where $\delta$ refers to a small constant to avoid division by zero.

\section{Experiments}
\subsection{Dataset}
In the experiments, we use the filtered dataset from CodeSearchNet\cite{husain_codesearchnet_2019} instead of the dataset used by baselines. Because the public dataset provided by Gu et at. is processed version, which could not be used in our model. CodeSearchNet contains the Java functions collected from the open-source code repositories. The train set and test set have already been split for avoiding functions from the same projects showing in both sides.

However, some code snippets in this dataset have useless annotations, e.g., annotations with only parameter descriptions. Thus, we filter out the code snippets following this criterion: the first sentence of its annotation should be longer than two words. After filtering, we obtain a train set containing {\raise.17ex\hbox{$\scriptstyle\mathtt{\sim}$}}330k annotation-function pairs and a test set containing {\raise.17ex\hbox{$\scriptstyle\mathtt{\sim}$}}19k annotation-function pairs. The detailed statistics are shown in Table~\ref{statistics}

\begin{table}[htbp]
\caption{Statistics of the datasets used in experiments}
\renewcommand{\arraystretch}{1.3}
\begin{center}
\begin{tabular}{lc}
\hline
\textbf{Statistics}\\
\hline
\# Train & 329,967 \\
\# Test & 19,015 \\
\hline
Average paths in code & 118.7 \\
Max paths in code & 500.0 \\
95th percentile paths in code& 369.0 \\
Average words in query & 10.2 \\
Max words in query & 440.0 \\
95th percentile words in query & 21.0 \\
\hline
\end{tabular}
\label{statistics}
\end{center}
\end{table}

With the filtered dataset, we perform the processing steps introduced in Section 3. Considering the scale of the model, we limit the size of the AST path, i.e., the maximum height and width are set to 8 and 3, respectively. The trained model searchs through a candidate pool consisting of all the code snippets in test set (i.e., 19,015 code snippets).

\subsection{Implementation Details}

We implement PSCS relying on several third-party libraries and frameworks.\footnote{Our implementation with data processing scripts for reproducing our model is available at https://github.com/v587su/PSCS\_public.} For data processing, we parse the function-level source code of Java and extract AST paths using PathMiner\cite{kovalenko_pathminer_2019}, an open-source java library for mining path-based representations of code. Besides, the neural model is implemented on top of Pytorch\footnote{https://pytorch.org/}, a widely used deep learning framework. 

During training, we use the Adam optimization algorithm\cite{kingma2014adam} to minimize the pairwise ranking loss, and the detailed parameters are listed in Table~\ref{parameter}. All the experiments run on a GPU environment under the Linux operating system (Ubuntu 16.04.1 LTS) with 64G DDR4 memory and a GTX 2080Ti GPU.

\begin{table}[htbp]
\caption{Parameters of PSCS}
\renewcommand{\arraystretch}{1.3}
\begin{center}
\begin{tabular}{lc}
\hline
\textbf{Parameters for optimization} & ~ \\
\hline
learning rate & 1e-4  \\
batch size & 64  \\
\hline
\textbf{Parameters for model} & ~ \\
\hline
word embedding size & 128  \\
node embedding size & 128  \\
dropout rate & 0.25 \\
LSTM hidden size & 128  \\
number of path $g$ & 100 \\
margin of rank loss & 1 \\
maximum length of path $p$ & 12 \\
maximum length of query $q$ & 20\\
\hline
\end{tabular}
\label{parameter}
\end{center}
\end{table}

\subsection{Evaluation Metrics}
Similar to the previous works\cite{wan_multi-modal_2019}\cite{shuai_improving_2020}, we do not perform manual evaluation considering human bias for a fair comparison. To evaluate our model, we measure SuccessRate@k and MRR. We report these evaluation metrics for the top-k search results because users can not inspect all the results.
Intuitively, SuccessRate@k is the percentage of getting correct search result among all test queries, calculated by:
\begin{equation}
    SuccessRate@k = \frac{1}{N}\sum_{i=1}^{N}isFound(S_i)
\end{equation}
and 
\begin{equation}
isFound(x) = \begin{cases}
    1 &\mbox{one result in $x$ is correct}\\
    0 &\mbox{otherwise}\\
\end{cases}
\end{equation}
where $N$ donates the size of the query set and $S_i$ is the search results for the $i$-th query. 

Compared to SuccessRate@k, MRR takes the rank of correct results into account. It is the average of the reciprocal ranks of results for test queries, calculated by:

\begin{equation}
    MRR = \frac{1}{N}\sum_{i=1}^{N}\frac{1}{rank_i}
\end{equation}

where $rank_i$ is the rank position of the correct code snippet in the result list for the $i$-th query.

In this experiment, we assume two cases, $k=1$ and $k=10$, which means that users only inspect the first result, and users inspect up to 10 results, respectively. 

\subsection{Baseline Models}
We implement and compare our model with two baseline models: DeepCS and CARLCS.

\textbf{DeepCS}\cite{gu_deep_2018}, as described in Section 2, is the first to adopt deep learning in code search task. It uses LSTM and max-pooling to encode the code features and queries. Its experimental results show that DeepCS outperforms the information-retrieval-based models.

\textbf{CARLCS}\cite{shuai_improving_2020}, recently proposed by Shuai et al., is the state-of-the-art model using a co-attention mechanism. A co-attention mechanism is introduced in CARLCS to learn the internal semantic correlations between code tokens and query tokens. In this work, we compare with its best-performing variant, CARLCS-CNN.

The public dataset used by these baseline models is the processed version, from which we can not extract the AST path. Thus, we reproduce the scripts for code processing and construct these models using their public code over the CodeSearchNet dataset. We use their default parameters except for the maximum length of queries. The distributions of query length are different in each dataset. For a fair comparison, we set this parameter to 20, which is more appropriate in our dataset.
\section{Results}

Our evaluation over Java code corpus focuses on the following research questions:

\textbf{RQ1:} How effective is PSCS at searching code snippets given queries in natural language?

\textbf{RQ2:} What is the contribution of code structure to the model effectiveness?

\textbf{RQ3:} What is the impact of each model component on the effectiveness of PSCS?

\textbf{RQ4:} How efficient is PSCS at searching code snippets given queries in natural language?

\textbf{RQ5:} How does the performance of PSCS change as the query length grows?

We address RQ1 to evaluate how effective PSCS is when compared to the existing state-of-the-art approaches. Then, we address RQ2 to provide insights into the importance of structural information. To the best of our knowledge, this is the first experiment to measure the impact of the pure code structure for semantic code search. RQ3 is addressed to understand the effectiveness of the alternative designs applied in this work. We address RQ4 to evaluate the efficiency of our approach. Lastly, we evaluate the sensitivity of PSCS to the length of input queries.

\subsection{RQ1 (Model Effectiveness): PSCS significantly outperforms all the baselines with a SuccessRate@10 of 47.6}

\begin{figure}[htbp]
\centerline{\includegraphics[width=9cm]{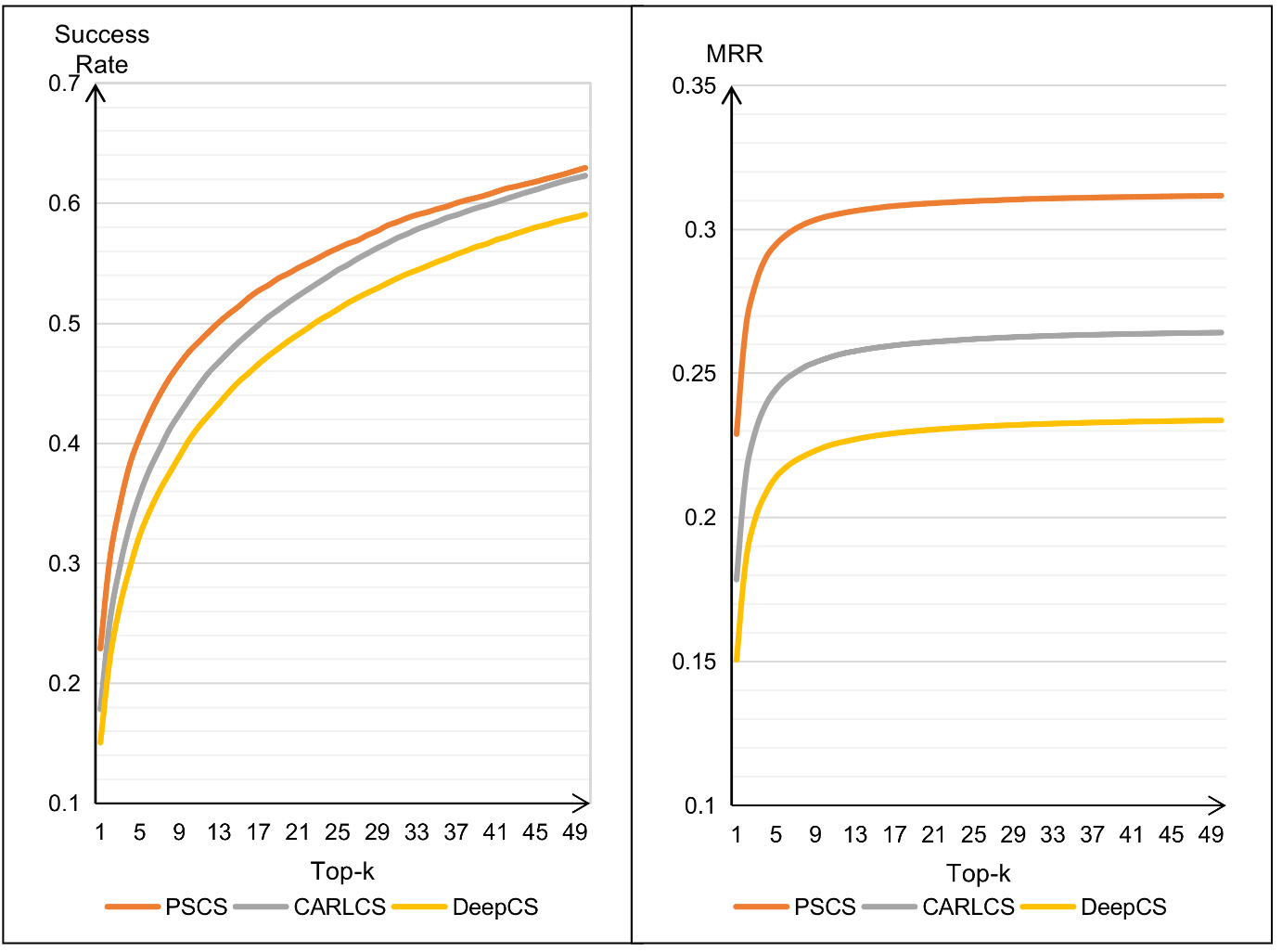}}
\caption{The performance changes of each model as the result list expands}
\label{vs_tokp}
\end{figure}

\begin{table}[htbp]
\caption{Evaluation results}
\renewcommand{\arraystretch}{1.3}
\begin{center}
\begin{tabular}{ccccc}
\hline
\textbf{Model} &\multicolumn{2}{c}{\textbf{SuccessRate}}&~&\multicolumn{1}{c}{\textbf{MRR}} \\
\cline{2-3}
\cline{5-5}
& \textbf{Top-1} & \textbf{Top-10} & ~ & \textbf{Top-10} \\
\hline
DeepCS & 14.6 & 40.3 &~ & 22.4 \\
CARLCS & 17.8 & 43.7 &~ & 25.5 \\
PSCS & \textbf{22.9} & \textbf{47.6} &~ & \textbf{30.4}\\
\hline
\end{tabular}
\label{effect_result}
\end{center}
\end{table}

With the experimental results, we compare the performance of PSCS with two baseline models, i.e., DeepCS and CARLCS. Table~\ref{effect_result} shows the SuccessRate and MRR of the code search results of each model, demonstrating that PSCS provides an improvement in model performance. It achieves 22.9\% SuccessRate@1. The figure increases to 47.6\% when top-10 results are considered, with an MRR of 30.4\%. We outperform all the baselines by a substantial margin, obtaining 19.2\% improvement in terms of MRR over the prior state of the art. Fig.~\ref{vs_tokp} shows the performance changes of each model as the size of the result list grows. As the size increases, all the models show a growth trend, and get stable gradually. As seen, we achieve the highest score among baselines across all result list sizes. 
These experimental results suggest that our model can more accurately match the related code snippets given a query. 

\subsection{RQ2 (The Impact of Code Structure): Code structure is an impactful supplementary feature for code understanding, contributing to a 38.8\% improvement in MRR}

\begin{table}[htbp]
\caption{Ablation study on the features of code}
\renewcommand{\arraystretch}{1.3}
\begin{center}
\begin{tabular}{lccccc}
\hline
\textbf{Ablation study} &\multicolumn{2}{c}{\textbf{SuccessRate}}&~&\multicolumn{2}{c}{\textbf{MRR}} \\
\cline{2-3}
\cline{5-6}
& \textbf{Top-1} & \textbf{Top-10} & ~ & \textbf{Top-10} & $\Delta$ \\
\hline
PSCS & \textbf{22.9} & \textbf{47.6} &~ & \textbf{30.4} & 0.0\\
\hline
- tokens & 0.1 & 1.0 &~ & 0.3 & -30.1\\
- nodes & 16.5 & 36.9 &~ & 21.9 & -8.5 \\

\hline
\end{tabular}
\label{path_result}
\end{center}
\end{table}

To measure the contribution of the code structure, we conducted an ablation study by removing features of AST paths.
In previous works\cite{wan_multi-modal_2019}\cite{shuai_improving_2020}, their code features hide the code structure in semantic tokens or hybrid tree nodes. They can not perform ablation studies on data only with code structure. In contrast, the AST path separate code structure and code semantics clearly. Hence, this is a good chance to evaluate the impact of the pure code structure:

\textbf{Tokens only}: We removed the non-terminal nodes in the AST path, only using terminal tokens to represent the code snippet. It means that the model needs to reason about code only with semantic information.

\textbf{Nodes only}: Different from \textit{Tokens only} variations, we hide the tokens and keep the nodes to evaluate the contribution of nodes in AST paths. The model has to understand the code only with structural information.

The results in Table~\ref{path_result} are quite impressive. Without tokens, \textit{nodes only} achieves an almost useless performance. It is comprehensible because even an experienced programmer can not understand a code snippet without any semantics (e.g., identifier names, API names, constant values). The MRR drops from 30.4\% to 21.9\% when the model only takes lexical tokens as input. Therefore, the contribution of the structural information from non-terminal nodes makes up 38.8\%. This ablation demonstrates the importance of code structure for semantic code search. Code structure used alone leads to useless results, but when combined with code semantics, a performance boost comes true. In conclusion, the code structure is an impactful supplementary feature for code understanding in code search. This conclusion provides suggestions on model design in the future.

\begin{figure*}[htbp]
\centerline{\includegraphics[width=13cm]{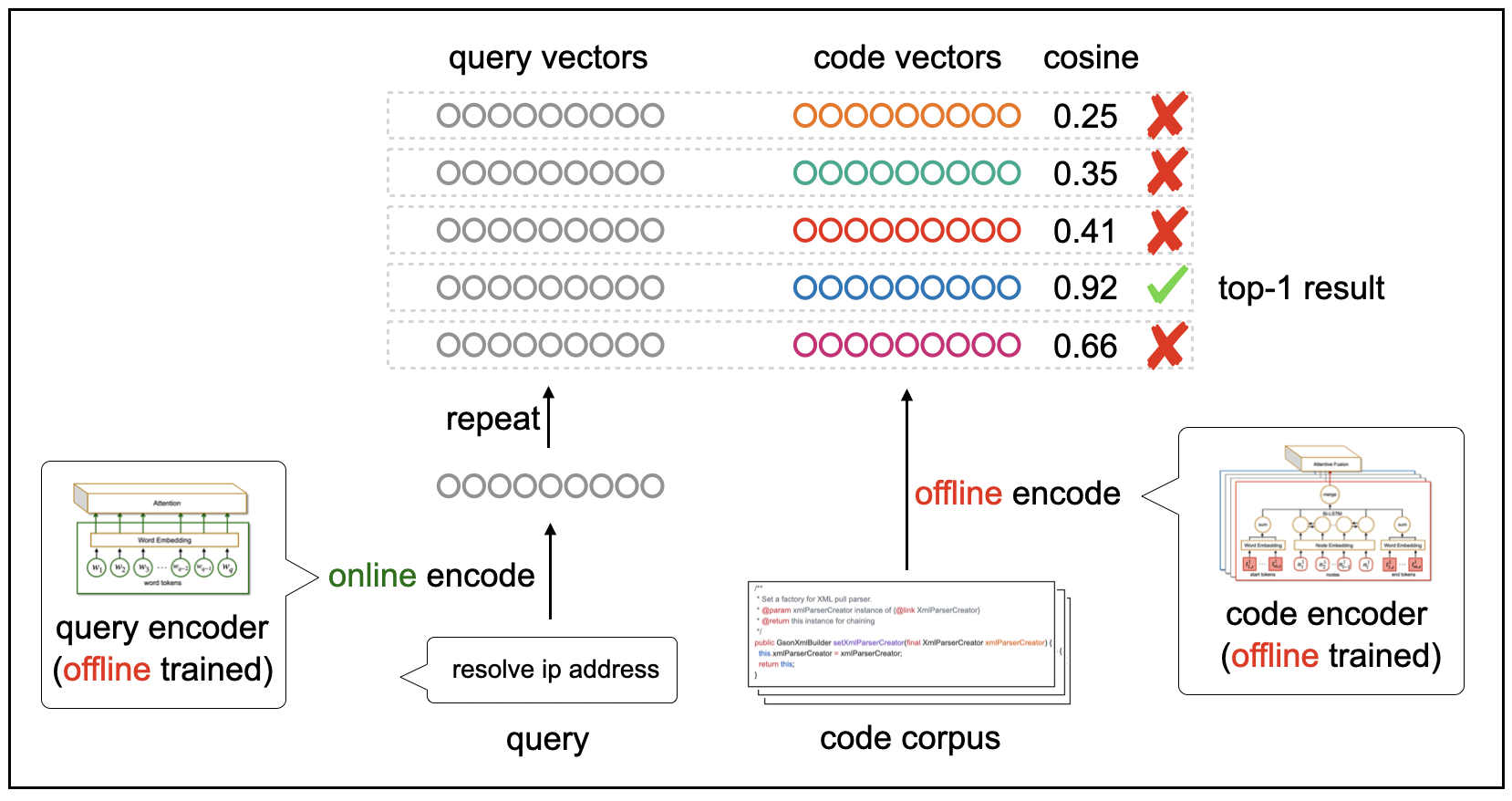}}
\caption{The search process of PSCS}
\label{search_logit}
\end{figure*}

\subsection{RQ3 (The Impact of Components): All the model components can boost total performance, especially the code attention and shared embedding}

\begin{table}[htbp]
\caption{Ablation study on model components}
\renewcommand{\arraystretch}{1.3}
\begin{center}
\begin{tabular}{lccccc}
\hline
\textbf{Ablation study} &\multicolumn{2}{c}{\textbf{SuccessRate}}&~&\multicolumn{2}{c}{\textbf{MRR}} \\
\cline{2-3}
\cline{5-6}
& \textbf{Top-1} & \textbf{Top-10} & ~ & \textbf{Top-10} & $\Delta$ \\
\hline
PSCS & \textbf{22.9} & \textbf{47.6} &~ & \textbf{30.4} & 0.0\\
\hline
- code attention & 11.5 & 29.9 &~ & 16.5 & -13.9 \\
- query attention & 21.5 & 45.0 &~ & 28.5 & -1.9\\
- shared embedding & 14.0 & 35.1 &~ & 20.0 & -10.4\\
- bi-LSTM & 19.6 & 42.2 &~ & 26.3 & -4.1 \\

\hline
\end{tabular}
\label{component_result}
\end{center}
\end{table}

To understand the contribution of each component, we experiment with removing several alternative designs:

\textbf{No code attention}: The code attention mechanism measures the contribution of each AST path. We removed this mechanism in the code encoder to evaluate the performance of code attention. The representation vectors of each path are averaged directly without attention weights.

\textbf{No query attention}: The query attention mechanism learns to measure the contribution of each word in a query. Similar to \textit{no code attention}, we use the average of word vectors to replace the query attention in query encoder.

\textbf{No shared embedding}: The shared embedding can help to narrow the semantic gap between query semantics and code semantics. In this ablation, we use different embedding matrices to embed the tokens in queries and terminals. 

\textbf{No bi-LSTM}: The bi-LSTM of code encoder is used to capture the sequential information from AST paths. In this ablation, we hide the bi-LSTM to measure the importance of the sequential information.

Table~\ref{component_result} demonstrates the decrement on metrics as the alternative design changes. \textit{No code attention} results in a drastic decrease of MRR (from 30.4\% to 16.5\%). This ablation shows that not all the paths in a code snippet are useful. An attentive weighted average can help to filter meaningless paths. In contrast, \textit{no code attention} only achieves a slight degradation, which means each word in a query sentence almost makes an equal contribution. As for \textit{no shared embedding}, using different embedding for query tokens and terminal tokens reduces MRR by more than one-third. It shows that the shared embedding works well in the code search task. Using a shared embedding can save the cost of learning. We believe this trick can be applied in further code search studies. The \textit{no bi-LSTM} ablation demonstrates the importance of sequential information in AST paths. Without sequential information, the performance decreases from 30.4\% to 26.3\% in terms of MRR.

\subsection{RQ4 (Model Efficiency): It takes 6ms for PSCS to handle a query to a code corpus containing over 19k code snippets}

\begin{table}[htbp]
\caption{The time cost of PSCS}
\renewcommand{\arraystretch}{1.3}
\begin{center}
\begin{tabular}{ll}
\hline
\textbf{Step} & \textbf{Time} \\
\hline
model training (offline) & 16 hours \\
code encoding (offline) & 3 ms / function \\
code search (online) & 6 ms / query \\
\hline
\end{tabular}
\label{time_cost}
\end{center}
\end{table}
As shown in Fig.~\ref{search_logit}, the time cost of our model consists of three parts: model training (offline), code encoding (offline), and code search (online). The figure indicates that the code vectors can be pre-encoded into vectors before user performing online search. From the aspect of users, the time cost of query encoding and similarity matching is the only problem they care about. 

We calculate the time per step to measure the efficiency of our model during the evaluation of PSCS over the test set. The results are shown in Table~\ref{time_cost}. The experiments run on a GTX 2080Ti GPU. The training time refers to the time from the beginning of training to the epoch that our model achieves its best performance. Model training is the most time-consuming part. Luckily, we do not need to re-train the model as long as the code corpus stays unchanged. 

As a deep learning model, the time cost for code search is usually lower than keyword-matching-based techniques. \cite{yan_are_2020}. It takes PSCS only 6ms to handle a query to the code corpus of the test set on average, which is acceptable in production environment. 

\subsection{RQ5 (The Sensitivity to query length): PSCS gives its best results for queries with 5 to 9 words}

\begin{figure*}[htbp]
\centerline{\includegraphics[width=11cm]{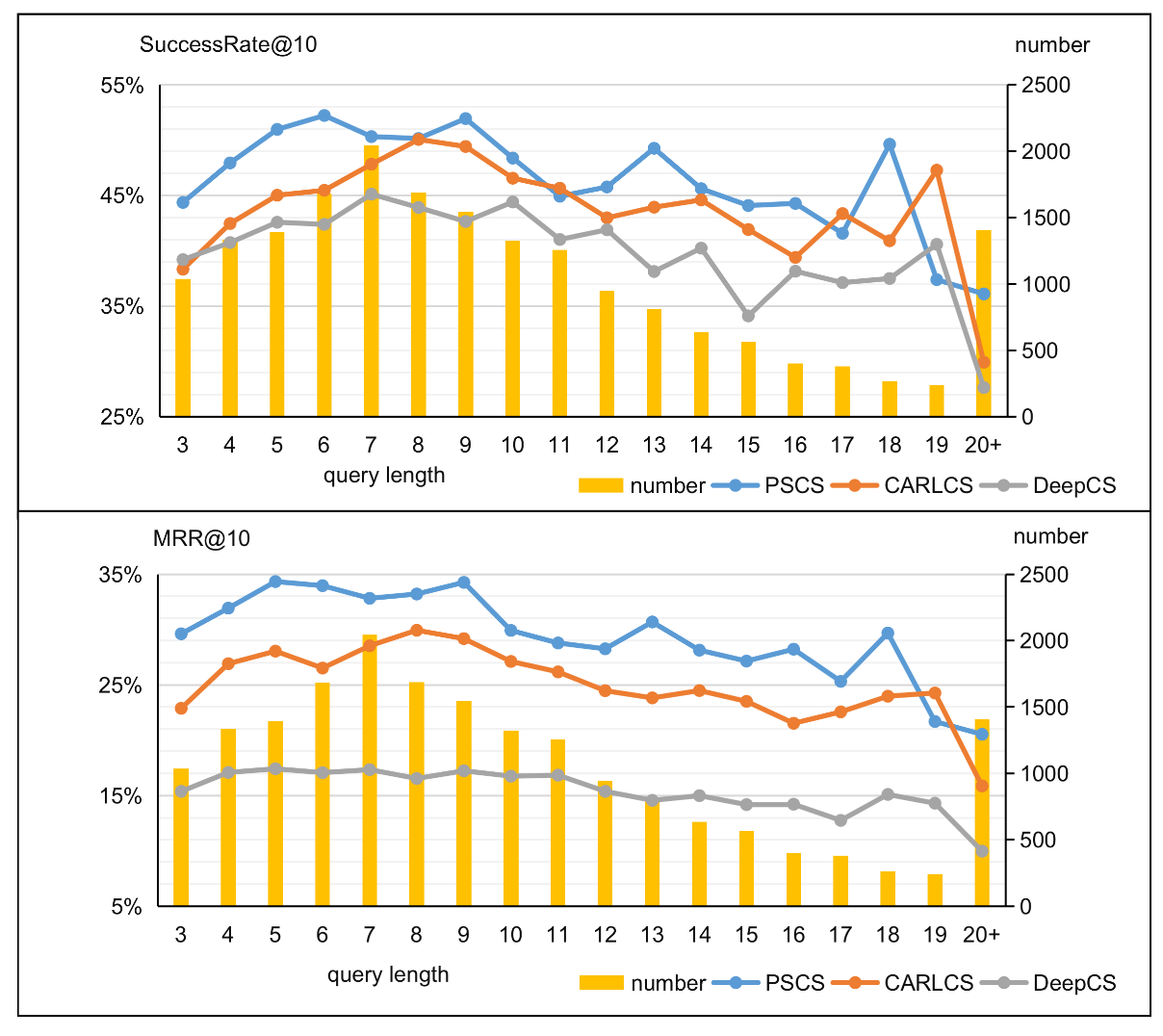}}
\caption{The change of SuccessRate@10 and MRR of each model as query length grows }
\label{query_length}
\end{figure*}

As shown in Fig.~\ref{query_length}, the SuccessRate@10 of PSCS is higher than 50\% when the query has 5 to 9 words. The most frequent length of query (2,043 queries have 7 words) also lies in this range, which means that our model can achieve its best performance with typical query length. Our model is superior to both baselines across most of the query lengths. As the query length increases, the performances of all models decrease gradually because of the additional noise from long queries. Besides, queries with few words may not contain the essential information for a model to identify the matched functions.

\subsection{Threats to Validity:}
We have identified the following threats to validity:

\textbf{A trade-off in the appropriateness of queries}: In this work, we take the first sentence of code annotation as the query. However, not all annotations are well-written. Most of the annotations are used within their projects, which limits the generalizability of queries. As a large-scale dataset, it is hard to correct the extracted queries manually. In recent work, researchers utilize a dataset with  Stack Overflow\footnote{https://stackoverflow.com/} question titles and code snippets answers. The question titles (usually start with "how to") are more query-appropriate than code annotations. Nevertheless, as resourceful as Stack Overflow is, the scale of questions is not large enough for a qualified dataset to train a deep learning model. Additionally, the code snippets extracted from answers are usually short and fragmentary.

\textbf{Baseline reproduction}: The baseline models use the same dataset collected by Gu et al.\cite{gu_deep_2018}. This dataset is a processed version that is incompatible with our model. Therefore, we perform experiments over a new dataset based on the open-source code of these two baselines. However, the scripts for data processing is not included in their public code. We have to process the data following the instructions in their original papers, which will introduce some slight bias caused by different third-party libraries and implementation details.

\textbf{A trade-off in automated evaluation}:
In practice, there are multiple related code snippets for a query. However, for automated evaluation, each query has limited correct code snippets. During the evaluation, the other results, which are also related, can not be recognized unless human involves. In particular, it is meaningless for our experiments to use metrics like Recall@k, Precision@k, and Normalized Discounted Cumulative Gain (NDCG) because these metrics are effective with more than one ground truth results. The lack of evaluation metrics may lead to an incomprehensive evaluation of our model.

\section{Conclusion}
In conclusion, we propose a path-based neural model for semantic code search to addresses the lack of structural information in code understanding. In contrast to previous works which relearn the structural information from code tokens, we represent code with the AST path. It consists of terminal nodes and non-terminal nodes representing the code structure and semantics, respectively. In this way, the structural information is directly fed into 
the neural model. Our model is designed to fit the AST path with a shared word embedding, a bi-directional LSTM model, and an attention mechanism to measure the contribution of each path dynamically. The experiments demonstrate that PSCS achieves a SuccessRate of 47.6\% and an MRR of 30.4\%, which significantly outperforms the state-of-the-art baselines. Besides, we are the first to demonstrate the importance of code structure for code search task. The experimental results of the ablation studies provide suggestions on model design for further studies. In the future, we plan to introduce more domain knowledge like programming syntax to promote the understanding of code structure for the code search task.

\bibliographystyle{ieeetr}

\begin{thebibliography}{10}

\bibitem{stolee_solving_2014}
K.~T. Stolee, S.~Elbaum, and D.~Dobos, ``Solving the {Search} for {Source}
  {Code},'' {\em ACM Transactions on Software Engineering and Methodology},
  vol.~23, pp.~1--45, June 2014.

\bibitem{lv_codehow_2015}
F.~Lv, H.~Zhang, J.-g. Lou, S.~Wang, D.~Zhang, and J.~Zhao, ``{CodeHow}:
  {Effective} {Code} {Search} {Based} on {API} {Understanding} and {Extended}
  {Boolean} {Model} ({E}),'' in {\em 2015 30th {IEEE}/{ACM} {International}
  {Conference} on {Automated} {Software} {Engineering} ({ASE})}, (Lincoln, NE,
  USA), pp.~260--270, IEEE, Nov. 2015.

\bibitem{raghothaman_swim_2016}
M.~Raghothaman, Y.~Wei, and Y.~Hamadi, ``{SWIM}: synthesizing what i mean: code
  search and idiomatic snippet synthesis,'' in {\em Proceedings of the 38th
  {International} {Conference} on {Software} {Engineering} - {ICSE} '16},
  (Austin, Texas), pp.~357--367, ACM Press, 2016.

\bibitem{niu_learning_2017}
H.~Niu, I.~Keivanloo, and Y.~Zou, ``Learning to rank code examples for code
  search engines,'' {\em Empirical Software Engineering}, vol.~22,
  pp.~259--291, Feb. 2017.

\bibitem{zhang_expanding_2018}
F.~Zhang, H.~Niu, I.~Keivanloo, and Y.~Zou, ``Expanding {Queries} for {Code}
  {Search} {Using} {Semantically} {Related} {API} {Class}-names,'' {\em IEEE
  Transactions on Software Engineering}, vol.~44, pp.~1070--1082, Nov. 2018.

\bibitem{nie_query_2016}
L.~Nie, H.~Jiang, Z.~Ren, Z.~Sun, and X.~Li, ``Query {Expansion} {Based} on
  {Crowd} {Knowledge} for {Code} {Search},'' {\em IEEE Transactions on Services
  Computing}, vol.~9, pp.~771--783, Sept. 2016.
\newblock arXiv: 1703.01443.

\bibitem{gu_deep_2018}
X.~Gu, H.~Zhang, and S.~Kim, ``Deep code search,'' in {\em Proceedings of the
  40th {International} {Conference} on {Software} {Engineering} - {ICSE} '18},
  (Gothenburg, Sweden), pp.~933--944, ACM Press, 2018.

\bibitem{hu2018deep}
X.~Hu, G.~Li, X.~Xia, D.~Lo, and Z.~Jin, ``Deep code comment generation,'' in
  {\em Proceedings of the 26th Conference on Program Comprehension},
  pp.~200--210, 2018.

\bibitem{sun_req2lib_2020}
Z.~Sun, Y.~Liu, Z.~Cheng, C.~Yang, and P.~Che, ``{Req2Lib}: {A} {Semantic}
  {Neural} {Model} for {Software} {Library} {Recommendation},'' in {\em 2020
  {IEEE} 27th {International} {Conference} on {Software} {Analysis},
  {Evolution} and {Reengineering} ({SANER})}, pp.~542--546, Feb. 2020.
\newblock ISSN: 1534-5351.

\bibitem{cambronero_when_2019}
J.~Cambronero, H.~Li, S.~Kim, K.~Sen, and S.~Chandra, ``When {Deep} {Learning}
  {Met} {Code} {Search},'' {\em arXiv:1905.03813 [cs]}, Oct. 2019.
\newblock arXiv: 1905.03813.

\bibitem{wan_multi-modal_2019}
Y.~Wan, J.~Shu, Y.~Sui, G.~Xu, Z.~Zhao, J.~Wu, and P.~S. Yu, ``Multi-{Modal}
  {Attention} {Network} {Learning} for {Semantic} {Source} {Code}
  {Retrieval},'' {\em arXiv:1909.13516 [cs]}, Sept. 2019.
\newblock arXiv: 1909.13516.

\bibitem{shuai_improving_2020}
J.~Shuai, L.~Xu, C.~Liu, M.~Yan, X.~Xia, and Y.~Lei, ``Improving {Code}
  {Search} with {Co}-{Attentive} {Representation} {Learning},'' {\em Accepted
  by 28th International Conference on Program Comprehension (ICPC)}, Oct. 2020.

\bibitem{alon_general_2018}
U.~Alon, M.~Zilberstein, O.~Levy, and E.~Yahav, ``A {General} {Path}-{Based}
  {Representation} for {Predicting} {Program} {Properties},'' {\em
  arXiv:1803.09544 [cs]}, Apr. 2018.
\newblock arXiv: 1803.09544.

\bibitem{husain_codesearchnet_2019}
H.~Husain, H.-H. Wu, T.~Gazit, M.~Allamanis, and M.~Brockschmidt,
  ``{CodeSearchNet} {Challenge}: {Evaluating} the {State} of {Semantic} {Code}
  {Search},'' {\em arXiv:1909.09436 [cs, stat]}, Sept. 2019.
\newblock arXiv: 1909.09436.

\bibitem{linstead_sourcerer_2009}
E.~Linstead, S.~Bajracharya, T.~Ngo, P.~Rigor, C.~Lopes, and P.~Baldi,
  ``Sourcerer: mining and searching internet-scale software repositories,''
  {\em Data Mining and Knowledge Discovery}, vol.~18, pp.~300--336, Apr. 2009.

\bibitem{chechik_sniff_2009}
S.~Chatterjee, S.~Juvekar, and K.~Sen, ``{SNIFF}: {A} {Search} {Engine} for
  {Java} {Using} {Free}-{Form} {Queries},'' in {\em Fundamental {Approaches} to
  {Software} {Engineering}} (M.~Chechik and M.~Wirsing, eds.), vol.~5503,
  pp.~385--400, Berlin, Heidelberg: Springer Berlin Heidelberg, 2009.
\newblock Series Title: Lecture Notes in Computer Science.

\bibitem{chen_neural_2018}
Q.~Chen and M.~Zhou, ``A neural framework for retrieval and summarization of
  source code,'' in {\em Proceedings of the 33rd {ACM}/{IEEE} {International}
  {Conference} on {Automated} {Software} {Engineering} - {ASE} 2018},
  (Montpellier, France), pp.~826--831, ACM Press, 2018.

\bibitem{ye_leveraging_2020}
W.~Ye, R.~Xie, J.~Zhang, T.~Hu, X.~Wang, and S.~Zhang, ``Leveraging {Code}
  {Generation} to {Improve} {Code} {Retrieval} and {Summarization} via {Dual}
  {Learning},'' p.~11, 2020.

\bibitem{yao_coacor_2019}
Z.~Yao, J.~R. Peddamail, and H.~Sun, ``{CoaCor}: {Code} {Annotation} for {Code}
  {Retrieval} with {Reinforcement} {Learning},'' in {\em The {World} {Wide}
  {Web} {Conference} on - {WWW} '19}, (San Francisco, CA, USA), pp.~2203--2214,
  ACM Press, 2019.

\bibitem{Wang2020TranS3AT}
W.-H. Wang, Y.~Zhang, Z.~Zeng, and G.~Xu, ``Trans$\hat{}$3: A transformer-based
  framework for unifying code summarization and code search,'' {\em ArXiv},
  vol.~abs/2003.03238, 2020.

\bibitem{loyola2017neural}
P.~Loyola, E.~Marrese-Taylor, and Y.~Matsuo, ``A neural architecture for
  generating natural language descriptions from source code changes,'' {\em
  arXiv preprint arXiv:1704.04856}, 2017.

\bibitem{bielik_phog_nodate}
P.~Bielik and V.~R. chev, ``Phog: Probabilistic model for code,'' in {\em
  ICML}, 2016.

\bibitem{murali2017neural}
V.~Murali, L.~Qi, S.~Chaudhuri, and C.~Jermaine, ``Neural sketch learning for
  conditional program generation,'' {\em arXiv preprint arXiv:1703.05698},
  2017.

\bibitem{mou2016convolutional}
L.~Mou, G.~Li, L.~Zhang, T.~Wang, and Z.~Jin, ``Convolutional neural networks
  over tree structures for programming language processing,'' in {\em Thirtieth
  AAAI Conference on Artificial Intelligence}, 2016.

\bibitem{allamanis_learning_2018}
M.~Allamanis, M.~Brockschmidt, and M.~Khademi, ``Learning to {Represent}
  {Programs} with {Graphs},'' {\em arXiv:1711.00740 [cs]}, May 2018.
\newblock arXiv: 1711.00740.

\bibitem{piech2015learning}
C.~Piech, J.~Huang, A.~Nguyen, M.~Phulsuksombati, M.~Sahami, and L.~J. Guibas,
  ``Learning program embeddings to propagate feedback on student code,'' {\em
  arXiv: Learning}, 2015.

\bibitem{alon_code2vec_2019}
U.~Alon, M.~Zilberstein, O.~Levy, and E.~Yahav, ``code2vec: learning
  distributed representations of code,'' {\em Proceedings of the ACM on
  Programming Languages}, vol.~3, pp.~1--29, Jan. 2019.

\bibitem{alon_code2seq_2019}
U.~Alon, S.~Brody, O.~Levy, and E.~Yahav, ``code2seq: Generating sequences from
  structured representations of code,'' {\em arXiv preprint arXiv:1808.01400},
  2018.

\bibitem{dam2018deep}
H.~K. Dam, T.~Pham, S.~W. Ng, T.~Tran, J.~Grundy, A.~Ghose, T.~Kim, and C.-J.
  Kim, ``A deep tree-based model for software defect prediction,'' {\em arXiv
  preprint arXiv:1802.00921}, 2018.

\bibitem{sun2019grammar}
Z.~Sun, Q.~Zhu, L.~Mou, Y.~Xiong, G.~Li, and L.~Zhang, ``A grammar-based
  structural cnn decoder for code generation,'' in {\em Proceedings of the AAAI
  Conference on Artificial Intelligence}, vol.~33, pp.~7055--7062, 2019.

\bibitem{hochreiter1997long}
S.~Hochreiter and J.~Schmidhuber, ``Long short-term memory,'' {\em Neural
  computation}, vol.~9, no.~8, pp.~1735--1780, 1997.

\bibitem{Srivastava2014DropoutAS}
N.~Srivastava, G.~E. Hinton, A.~Krizhevsky, I.~Sutskever, and R.~Salakhutdinov,
  ``Dropout: a simple way to prevent neural networks from overfitting,'' {\em
  J. Mach. Learn. Res.}, vol.~15, pp.~1929--1958, 2014.

\bibitem{bahdanau2014neural}
D.~Bahdanau, K.~Cho, and Y.~Bengio, ``Neural machine translation by jointly
  learning to align and translate,'' {\em arXiv preprint arXiv:1409.0473},
  2014.

\bibitem{kovalenko_pathminer_2019}
V.~Kovalenko, E.~Bogomolov, T.~Bryksin, and A.~Bacchelli, ``{PathMiner}: {A}
  {Library} for {Mining} of {Path}-{Based} {Representations} of {Code},'' in
  {\em 2019 {IEEE}/{ACM} 16th {International} {Conference} on {Mining}
  {Software} {Repositories} ({MSR})}, (Montreal, QC, Canada), pp.~13--17, IEEE,
  May 2019.

\bibitem{kingma2014adam}
D.~P. Kingma and J.~Ba, ``Adam: A method for stochastic optimization,'' {\em
  arXiv preprint arXiv:1412.6980}, 2014.

\bibitem{yan_are_2020}
S.~Yan, H.~Yu, Y.~Chen, B.~Shen, and L.~Jiang, ``Are the {Code} {Snippets}
  {What} {We} {Are} {Searching} for? {A} {Benchmark} and an {Empirical} {Study}
  on {Code} {Search} with {Natural}-{Language} {Queries},'' in {\em 2020 {IEEE}
  27th {International} {Conference} on {Software} {Analysis}, {Evolution} and
  {Reengineering} ({SANER})}, (London, ON, Canada), pp.~344--354, IEEE, Feb.
  2020.

\end{thebibliography}

\end{document}